\documentclass[fleqn,usenatbib]{mnras}

\usepackage{newtxtext,newtxmath}

\usepackage[T1]{fontenc}
\usepackage[justification=centering]{caption}
\DeclareRobustCommand{\VAN}[3]{#2}
\let\VANthebibliography\thebibliography
\def\thebibliography{\DeclareRobustCommand{\VAN}[3]{##3}\VANthebibliography}

\usepackage{graphicx}	
\usepackage{amsmath}	
\usepackage{orcidlink}

\newcommand{\Muv}{$\rm{M}_{UV}\,$}
\newcommand{\Mstar}{$\log{\rm{M}_*/\rm{M}_\odot}\,$}
\DeclareRobustCommand{\xiion}{$\xi_{\rm{ion}}\,$}	
\newcommand{\xiunit}{$\rm{Hz}\,\rm{erg}^{-1}$}	

\newcommand{\buv}{$\beta_{1500}\,$}

\title[Dwarf Star Forming Galaxies at z$\sim$5]{CANUCS: UV and Ionising Properties of Dwarf Star Forming Galaxies at z $\sim$ 5 to 7 }
\author[A. Harshan et al.]{
Anishya Harshan,$^{1}$\thanks{E-mail: anishya.thevalil-harshan@fmf.uni-lj.si}\orcidlink{0000-0001-9414-6382},
Maruša Bradač$^{1}$\orcidlink{0000-0001-5984-0395},
Roberto Abraham$^{2,3}$\orcidlink{0000-0002-4542-921X},
Yoshihisa Asada$^{4,5}$\orcidlink{0000-0003-3983-5438},
Gabriel Brammer$^{6}$\orcidlink{0000-0003-2680-005X},\newauthor
Guillaume Desprez$^{4}$\orcidlink{0000-0001-8325-1742},
Karthiek Iyer$^{8}$\orcidlink{0000-0001-9298-3523},
Nicholas S. Martis$^{1,4,9}$\orcidlink{0000-0003-3243-9969},
Jasleen Matharu$^{6,7}$\orcidlink{0000-0002-7547-3385},
Lamiya Mowla$^{10}$\orcidlink{0000-0002-8530-9765},\newauthor
Adam Muzzin$^{11}$,
Ga\"el Noirot$^{4}$,
Gregor Rihtaršič$^{1}$\orcidlink{0009-0009-4388-898X},
Ghassan T. E. Sarrouh$^{11}$\orcidlink{0000-0001-8830-2166},
Marcin Sawicki$^{4}$\orcidlink{0000-0002-7712-7857},\newauthor
Victoria Strait$^{6,7}$\orcidlink{0000-0002-6338-7295}, 
Chris J. Willott$^{9}$\orcidlink{0000-0002-4201-7367},
\\
$^{1}$ Department of Mathematics and Physics, Jadranska ulica 19, SI-1000 Ljubljana, Slovenia\\
$^{2}$ David A. Dunlap Department of Astronomy and Astrophysics, University of Toronto, 50 St. George Street, Toronto, Ontario, M5S 3H4, Canada\\
$^{3}$ Dunlap Institute for Astronomy and Astrophysics, 50 St. George Street, Toronto, Ontario, M5S 3H4, Canada\\
$^{4}$ Department of Astronomy \& Physics and Institute for Computational Astrophysics, Saint Mary's University, 923 Robie Street, Halifax, NS B3H 3C3, Canada\\
$^{5}$ Department of Astronomy, Kyoto University, Sakyo-ku, Kyoto 606-8502, Japan\\
$^{6}$ Niels Bohr Institute, University of Copenhagen, Jagtvej 128, DK-2200 Copenhagen N, Denmark\\
$^{7}$ Cosmic Dawn Center (DAWN), Denmark\\
$^{8}$Columbia Astrophysics Laboratory, Columbia University, 550 West 120th Street, New York, NY 10027, USA\\
$^{9}$ National Research Council of Canada, Herzberg Astronomy \& Astrophysics Research Centre, 5071 West Saanich Road, Victoria, BC, V9E 2E7, Canada\\
$^{10}$ Whitin Observatory, Department of Physics and Astronomy, Wellesley College, 106 Central Street, Wellesley, MA 02481, USA\\
$^{11}$ Department of Physics and Astronomy, York University, 4700 Keele St. Toronto, Ontario, M3J 1P3, Canada
}

\date{Accepted XXX. Received YYY; in original form ZZZ}

\pubyear{2024}

\begin{document}
\label{firstpage}
\pagerange{\pageref{firstpage}--\pageref{lastpage}}
\maketitle

\begin{abstract}
The epoch of reionisation progressed through the emission of ionising photons from galaxies to their local intergalactic medium. In this work, we characterise the dwarf star-forming galaxies as candidates for the source of ionising photons that drove EoR. We investigate the ionising properties and star formation histories of star-forming dwarf galaxies at the last stages of EoR at $4.8<\rm{z}<7$ using observations from the CAnadian NIRISS Unbiased Cluster Survey (CANUCS). The magnification due to gravitational lensing allows us to probe large dynamic ranges in stellar mass ($2\times 10^{6}\leq\rm{M}_*/\rm{M}_\odot\leq5\times 10^{9}$) and UV magnitudes ($-22.68\leq$M$_{UV}\leq=-15.95$).We find a median UV slope \buv of $-2.
56\pm0.23$ and the production efficiency of ionising photons $\log$ \xiion $=25.39\pm0.6$ for the full sample ($4.8<\rm{z}<7$) with a median stellar mass of $6.3\pm0.5\times10^{7} \rm{M}_\odot$. We find both \buv and \xiion are marginally correlated with the stellar mass of the galaxy, indicating a possible greater contribution of dwarf galaxies to the reionisation of the Universe. We find that on average, galaxies in our sample are experiencing a recent rise/burst of star formation which translates to a higher scatter in \xiion and a large scatter in H$\alpha$ equivalent widths. Finally, we investigate the trends of  H$\alpha$ and [OIII]+H$\beta$ EWs with UV magnitude and find M$_{UV}$ is correlated between H$\alpha$ but not with  [OIII]+H$\beta$ EWs indicating low metallicities and recent burst in the UV faint galaxies.
\end{abstract}




\section{Introduction}


The production of ionising photons is a fundamental process that plays a crucial role in shaping the reionisation history of the Universe. However, the sources of ionizing photons that drove reionisation at redshifts z $> 5.5$ (e.g., \citealp{Bolan2022, Davies2018, McGreer2015}) are poorly understood. Current observational evidence suggests that reionisation proceeded through the escape of ionizing photons from young massive stellar populations in galaxies (see review \cite{Robertson2022} and references within). To ascertain the contribution of different galaxies to reionisation, we need to determine the rate of ionizing photons emitted into the inter-galactic medium (IGM) from different galaxy populations. The rate of ionizing photons emitted by galaxies can be determined by combining the non-ionising UV luminosity function, escape fraction of ionising photons into the IGM and the ionising photon production efficiency. 

The ionising photon production efficiency (\xiion) is the measure of rate of hydrogen-ionising photon production ($\rm{E}\geq 13.6$ eV) by young massive stars in a galaxy per unit non-ionising UV continuum luminosity produced on average by the less massive stars. Thus \xiion should depend on the fraction of massive ionising stars which is regulated by the initial mass function and the star formation history of the galaxy. Along with the escape fraction of ionising photons ($f_{\rm{esc}}$), \xiion is used to determine whether a galaxy is capable of reionising the local IGM. Direct measurement of \xiion requires detection of ionising Ly-continuum photons emitted by the galaxy which becomes increasingly difficult for high redshift galaxies as the radiation would be absorbed by the intergalactic medium. Thus the measurement of \xiion for high redshift galaxies has exclusively relied on indirect methods. As the bright nebular recombination line H$\alpha$ (6562\AA) is directly related to the total number of Ly-continuum photons produced by stars in a galaxy (assuming an escape fraction of zero : \citealp{Leitherer1995}), H$\alpha$ line can be used to constrain \xiion (e.g., \cite{Bouwens2016, Chisholm2022, Stefanon2022,Gonzalo2023}).


The past decade has seen many studies estimating \xiion using indirect H$\alpha$ at z $\sim$ 4 to 5  based on stacking Spitzer/IRAC  photometry (e.g.,\cite{Bouwens2016, Lam2019, Stefanon2019}) and at lower redshifts z$\sim$1 to 3.7, where  H$\alpha$ or H$\beta$ was observable from ground based telescopes \citep{Nakajima2016, Nanayakkara2020,Matthee2017}. These studies indicate that the \xiion increases with redshift, and thus the average escape fraction required for galaxies to ionise the local IGM would be $\approx 10 - 20 \%$ \citep{Ouchi2009, Robertson2015, Finkelstein2019, Naidu2020} at z = 6. However, most studies relied on low statistics, lower redshift analogs or wide band photometry and image stacking due to lack of access to Balmer lines. This has been mitigated  with the launch of the James Webb Space Telescope (JWST), and we are able to observe Balmer lines using spectroscopy and/or medium/narrow band photometry to accurately measure H$\alpha$ fluxes. Recently a number of studies using JWST observations like \cite{Lin2023,simmonds2023, Tang2023, Saxena2023, Gonzalo2023,Atek2023,Mascia2023} have measured \xiion, however, these studies rely on luminous and/or high mass galaxies or small samples for dwarf galaxies. Thus the ionisation properties of faint, low-mass galaxies remain unclear.

Observational studies suggest that low-mass galaxies should have a high contribution to the reionisation of the Universe because of their abundance indicated by the steep slope at the faint end of the UV luminosity function in the high redshift Universe \citep{Sawicki2006, Reddy2009, Bouwens2012, Dressler2015, Finkelstein2015, Ishigaki2015, Livermore2017, Mehta2017, Atek2018, Bhatawdekar2018, Atek2023}. The same is also suggested by simulation studies like \cite{shreedhar2023}, who suggest that inclusion of low mass halos significantly increases the progression of reionisation. Additionally, owing to the low gravitational potential along with stochastic star formation histories, low stellar mass galaxies are thought to have a higher escape fraction of ionising photons into the local IGM \citep{Paardekooper2013, Wise2014, Erb2015,Anderson2017, Karman2017} at high redshifts. In order to determine the contribution of low-mass galaxies to the reionisation of the Universe, we need to investigate their star formation and ionising properties and compare them to their massive counterparts. However, due to the faintness of low-mass galaxies, most studies have focused on more massive and luminous galaxies.

To that end, we use the CAnadian NIRISS Unbiased Cluster Survey \citep[CANUCS]{willott2022}, to study the low mass galaxies at redshift z $>5$.  The magnification provided by the lensing galaxy clusters along with the capabilities of the JWST means we are now able to spectroscopically observe low stellar mass faint galaxies that were previously hidden at z $>5$. With CANUCS, we are able to reach UV magnitude \Muv $\approx -15.95$ compared to \Muv $\approx -16.4$ from the deep imaging from the JADES \citep{Eisenstein2023,Endsley2023} for galaxies at $6<\rm{z}<9$.  We use medium and broad band photometry from NIRCam and prism spectroscopy from NIRSpec to study the ionisation properties and star formation histories of star-forming galaxies at $5<\rm{z}<7$. The outline of the paper is as follows: We describe the observations, data reduction, and sample selection in section 2. In section 3, we describe the methodology followed for flux measurements from photometric and spectroscopic observations, dust correction, and measurements of UV and ionising properties. In section 4, we discuss the results on the UV slope, \xiion,  and emission line equivalent widths and finally summarise our findings in section 5.

We assume a ﬂat $\Lambda$CDM cosmology with $\Omega_{m} = 0.3, \Omega_{\Lambda} =0.7, \rm{h} = 0.7$, all magnitudes are in the AB system.

\section{DATA and Target Selection}

\subsection{Imaging Data}
In this work, we use data from the Canadian NIRISS Unbiased Cluster Survey \citep[CANUCS]{willott2022}, a JWST GTO program (Program ID 1208; PI C.Willott) and the HST program 16667 (PI M. Bradač). We select our sample from the observations of the MACS J0417.5-1154 cluster (CLU) field and the respective NIRCam flanking (NCF) field. In the central CLU field, the CANUCS imaging data consists of HST/ACS filters: F435W, F606W, F814W and JWST/NIRCam filters: F090W, F115W, F150W, F200W, F277W, F356W, F410M, F444W. In the NCF field we have HST/WFC3 filters: F438M, F606W and JWST/NIRCam filters: F090W, F115W, F140M, F150W, F162M, F182M, F210M, F250M, F277W, F300M, F335M, F360M, F410M, F444W. CANUCS image reduction and photometry procedure is described in detail in \cite{Noirot2023}. In short, we use a modified version of the Detector1Pipeline (calwebb\textunderscore detector1) stage of the official STScI pipeline and jwst\textunderscore 0916.pmap JWST Operational Pipeline (CRDS\textunderscore CTX) to reduce the NIRCam data. We perform astrometric alignment of the different exposures of JWST/NIRCam to HST/ACS images, sky subtraction, and drizzling to a common pixel scale of 0.04$''$ using version 1.6.0 of the grism redshift and line analysis software for space-based spectroscopy \citep[Grizli]{Brammer2021}. The source detection and photometry is done with the Photutils package \citep{Bradley2022} on the $\chi_{mean}$ detection image created using all available images. For each detected source, total fluxes are measured in elliptical apertures based on the Kron radius and circular apertures of diameter 0.3$\arcsec$ and 0.7$\arcsec$ \citep{Asada2024}. In this work, we use $0.7 \arcsec$-diameter aperture photometry on HST/ACS and JWST/NIRCam images that were PSF-homogenized to the resolution of the F444W data (Sarrouh et al., in prep). We calculate the photometric redshifts using EAZY-py \citep{Brammer2008,eazy2021} with SED templates from \cite{Larson2022} that are based on FSPS \citep{Conroy2010} and BPASS \citep{Eldridge2017} stellar population synthesis models and CLOUDY \citep{Ferland2017, cloudy2018} photoionization code. 

We select our photometric sample between redshift 4.8 and 5.5 such that the medium band F410M of JWST/NIRCam includes flux from the H$\alpha$ (6562\AA) emission line creating a flux excess. We get a sample of 270 objects with flux excess in F410M (i.e., $\rm{Flux}_{\rm{F410M}} > \rm{Flux}_{\rm{F444W}}$) in the selected redshift range. We impose SNR cuts for Lyman-break dropout ($\rm{SNR}_{\rm{F606W }}< 2\, \rm{and}\, \rm{SNR}_{\rm{F090W}} >3 $) and  visually inspect each object and select our final sample of 156 galaxies. We refer to this sample as the NIRCam sample throughout the paper.

\subsection{Spectroscopic Data}
Galaxies selected with excess flux in F410M and Lyman break drop-out in F606W, F814W or F090W were selected for spectroscopic follow-up without imposing a magnitude limit with JWST/NIRSpec micro-shutter assembly in the low-resolution Prism mode. Our NIRSpec data has been reduced using the JWST pipeline for stage 1 corrections and then the msaexp \citep{Brammer2022} to create wavelength calibrated, background subtracted 2D spectra. The 1D spectrum is then created for each line by collapsing the spectrum in the spatial axis and extracting the 1D spectrum within 3 $\sigma$ of the peak of the collapsed spectrum. From the CANUCS spectroscopic observations, we confirm a sample of 33 galaxies with H$\alpha$ emission line at $4.8<z<7$.

The wavelength coverage of the NIRSpec prism allows for observation of H$\alpha$ emission at $4.8<\rm{z}<7$. Thus, our NIRSpec data covers a higher redshift range compared to our NIRCam sample which covers $4.8<\rm{z}<5.5$. The NIRCam data is limited in wavelength to allow for accurate measurement of H$\alpha$ flux, which can be achieved when H$\alpha$ emission falls in the F410M band (in $4.8<\rm{z}<5.5$). The lower observational depth of the spectroscopic data compared to the NIRCam imaging leads to a sample of lower magnification uncorrected M$_{UV}$ of -24.37 to -18.45 with median M$_{UV}=  -20.95 \pm 1.15$ in our NIRSpec Sample compared to uncorrected the M$_{UV}$ range of -23.13 to -18.10 with median M$_{UV}= -20.56 \pm 0.85$ in our NIRCam Sample. Similarly, the magnification uncorrected stellar masses of the NIRSpec sample is in the range  7.78 $<$ \Mstar $<$ 9.51 with a median \Mstar $=8.11 \pm0.42$ compared to the \Mstar range 7.67 $<$ \Mstar $<$ 9.75 with a median \Mstar $=8.04 \pm0.38$ for the NIRCam sample. The properties of the two sample sets are presented in Table 1 and shown in Figure \ref{fig:hist}. 

\begin{table*}
\begin{center}
\captionsetup{justification=centering}

\label{tab:table}
    
\begin{tabular}{*{6}{c}}

 \hline
 Sample & z & $\log \rm{M}_*/\rm{M}_\odot$ & M$_{UV}$ & $\beta_{1500}$ & \xiion \\ [0.5ex] 
 \hline\hline
 \shortstack{NIRCam \\ \, }& \shortstack{5.15 $\pm$ 0.2 \\(4.8 - 5.5)} & \shortstack{7.8 $\pm$ 0.5 \\ (6.3 - 9.7)}  & \shortstack{ -20.0 $\pm$ 1.2\\(-22.7 - -15.9)} & \shortstack{ -2.58 $\pm$ 0.2\\(-3.15 - -0.83)} & \shortstack{25.36 $\pm$ 0.04 \\(24.26 - 26.37)}\\ 
 \hline
\shortstack{NIRSpec\\ \, } & \shortstack{5.5 $\pm$ 0.6 \\(4.8 - 7)} & \shortstack{7.9 $\pm$ 0.5\\(6.7 - 9.2)} & \shortstack{ -20.1 $\pm$ 1.2\\(-21.8 - 17.0)} & \shortstack{-2.48 $\pm$ 0.4 \\(-2.78 - -1.73 ) } & \shortstack{25.48 $\pm$ 1.0 \\(21.25 -  26.75)}\\ [1ex] 
 \hline

\end{tabular}
 \caption{Summary of median values and ranges (in parantheses) of physical properties of the NIRCam and NIRSpec sample. Stellar mass ($\log \rm{M}_*/\rm{M}_\odot$), UV magnitude (M$_{UV}$) are corrected for magnification. }
\end{center}
\end{table*}

\section{Methods}
\subsection{SED fitting and Flux Measurements}
\label{sec:sedfit}
\begin{figure*}
    
    \includegraphics[scale =0.6]{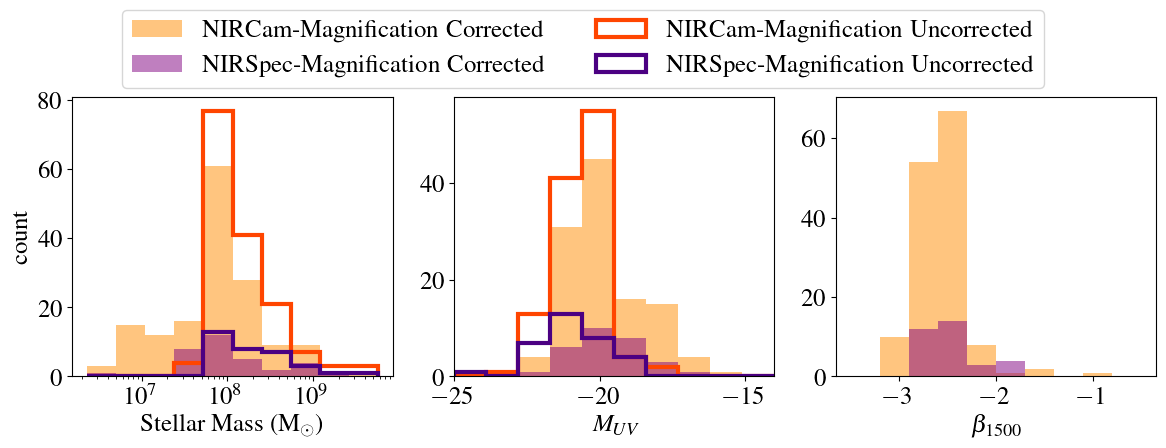}
    \caption{Distribution of Stellar mass (left), UV Magnitude in AB system (center) and the UV slope at 1500 \AA (right) for our selected sample. The orange histograms represent the NIRCam sample at redshift $4.8>\rm{z}>5.5$, the violet histograms represent the NIRSpec sample at $4.8>\rm{z}>7$. The filled and unfilled histograms represent the magnification corrected and uncorrected quantities respectively. Gravitational lensing correction improves our detection limits by 1 dex in \Mstar and 2.1 dex in \Muv. }
    \label{fig:hist}
\end{figure*}

We calculate the photometric redshifts using EAZY-Py \citep{Brammer2008,eazy2021} including SED templates from \cite{Larson2022} that are better suited to fit the SEDs of blue high redshift ($5<\rm{z}<8$) galaxies. We used Dense Basis \citep{ Iyer2017,Iyer2019} SED fitting code to determine other physical properties of galaxies: stellar masses, star formation rates (SFRs), metallicities, stellar dust attenuation and non-parametric star formation histories (SFHs). Dense Basis uses Flexible Stellar Population Synthesis (FSPS) code \citep{Conroy2010} which includes a prescription for nebular emission from CLOUDY \citep{Ferland2017,cloudy2018}. We adopt the \cite{Calzetti2000} dust model and assume \cite{Chabrier2003} IMF.  We fit the photometry from 3 HST/ACS bands and 8 JWST/NIRCam bands covering a wavelength range of 0.4 $\mu m$ to 4 $\mu m$ for the CLU field and 2 HST/WFC3 bands and 14 JWST/NIRCam bands for the NCF field covering a wavelength range of 0.4 $\mu m$ to 4 $\mu m$. The photometric sample is selected with H$\alpha$ flux excess in the F410M medium band and has been visually inspected to be a Lyman-continuum drop-out in the F606W or F814W bands. Thus, we run Dense Basis with uniform redshift prior in $4.7<\rm{z}<6$ range. For the spectroscopic sample, we fix the redshift at the measured spectroscopic redshift while running Dense Basis. We use flat stellar mass prior between $5<$\Mstar $<12$, flat sSFR prior between $-12<$ sSFR ($yr^{-1}$) $<-7$, varied metallicity in the range $-2.5 <\log(Z/Z_\odot)<0.25$ and used an exponential prior for dust attenuation. We then correct the stellar mass, SFR, and SFHs for magnification from gravitational lensing from the foreground cluster. The magnification coefficients are calculated using an updated lens model for MACS0417 (Desprez 2024, in prep.) created with Lenstool \citep{Kneib1993, Jullo2007}. The model improves upon the \cite{Mahler2019} model by adding several new spectroscopically confirmed multiply imaged systems. 

Figure \ref{fig:hist} shows the stellar mass distributions before and after magnification corrections. Our sample shows (not corrected for magnification) a lower limit of \Mstar $= 7.3$ (orange unfilled histogram, left) for the NIRCam sample and \Mstar $= 7.78$ (purple unfilled histogram, left) for the NIRSpec sample. After correcting for magnification from gravitational lensing from the cluster, the stellar mass sample ranges between $6.36<$ \Mstar $<9.76$ with median \Mstar $= 7.88\pm 0.57$ and $6.69<$ \Mstar $<9.21$ with median \Mstar $= 7.94\pm 0.53$ for the NIRCam and NIRSpec samples respectively shown in filled histograms in \ref{fig:hist}. In our sample, the median lensing magnification is $3\pm1$. Given the distance of the NCF field from the cluster center, we assume a magnification of 1.

We measure the H$\alpha$ flux from the photometry by following the two-filter method  and the three-filter method described in \cite{Vilella-Rojo2015} for the CLU and NCF field respectively. In the CLU field, we use the F444W broad band filter to describe the stellar continuum and the F410M medium band with flux excess from H$\alpha$ ($4.8<\rm{z}<5.5$). In the NCF field, we calculate the continuum with the F360M medium band filter and the F444W broad band filter and use the F410M medium band that contains the H$\alpha$ flux. Similarly, only in the NCF field, we calculate the [OIII]+H$\beta$ flux for the NIRCam sample using the F300M, F335M, and F360M bands. We do not calculate the [OIII]+ H$\beta$ flux for galaxies in the CLU field as the [OIII]+ H$\beta$ falls in the F277W and F356W broad bands for the NIRCam sample which would include spectral features like [OII], Balmer break, etc. We calculate the restframe equivalent width (EW) for each emission line measurement using the extracted emission line following \cite{Vilella-Rojo2015}. We bootstrap the photometry within the photometric errors and measure the line fluxes and restframe EWs. We take the median values and $1\sigma$ of the distribution as the measured line flux and EWs and associated errors.

\begin{figure}
    \includegraphics[scale =0.6]{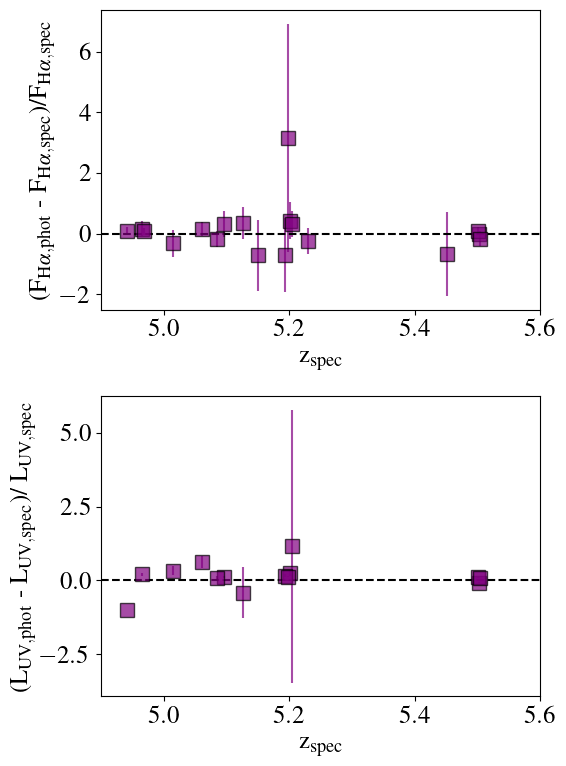}

    \caption{ Comparison of H$\alpha$ flux (top), and UV Luminosity (bottom) measured from photometry and spectroscopic data. Our photometric measurements of  H$\alpha$ flux (from F410M) and $\rm{L}_{UV}$ (from F090W) are consistent with the spectroscopic measurements.} 
    \label{fig:spec_phot}
\end{figure}

For our NIRSpec data, we use the ID spectra and correct for slit loss by performing a spline interpolation of the ratio of mock photometry from the 1D spectrum and the observed photometry and smoothing the interpolated function. For the spline interpolation, we also ignored the photometric bands where the mock photometry from the NIRSpec spectrum was less than 2 SNR. We then extract the 1D spectrum in 200 \AA\ wavelength windows around emission lines H$\alpha$, [OIII], and H$\beta$ and subtract the local continuum if present. Fluxes and accurate redshifts are then calculated by fitting Gaussian curves on the emission line without putting constraints on the width of the Gaussian. We bootstrap the spectrum with the error spectrum and calculate fluxes. The median flux and redshift and the $1\sigma$ of the distribution of the bootstrap are taken as the measured values and associated errors. Given the low resolution of the spectroscopic data from the NIRSpec prism, we note that the H$\alpha$ emission may have contamination from [NII](6585\AA) emission, and the [OIII] (5007,4959\AA) doublet is not always fully resolved (depending on SNR). \cite{Cameron2023} found no significant observation of [NII] in deep observation of Lyman break galaxies in $\rm{z} \sim 5.5 - 9.5 $. Thus we assume a negligible contribution of [NII] to the measured H$\alpha$ flux. We follow a similar assumption while calculating H$\alpha$ flux from the medium band photometry. However, we note that if we are to assume $\approx10\%$ contribution of [NII] as is observed at $z\sim2$ \citep{coil2015, alcorn2019}, our H$\alpha$ fluxes and quantities measured with H$\alpha$ fluxes will be lowered by $\approx10\%$. Figure \ref{fig:spec_phot} shows reasonable agreement ( $<7\%$ difference) between the H$\alpha$ flux measured with photometric data vs spectroscopy. We note that this comparison has been done within the overlapping NIRCam+NIRSpec sample (at $4.8<\rm{z}<5.5$).

\begin{figure}
    \includegraphics[scale=0.5]{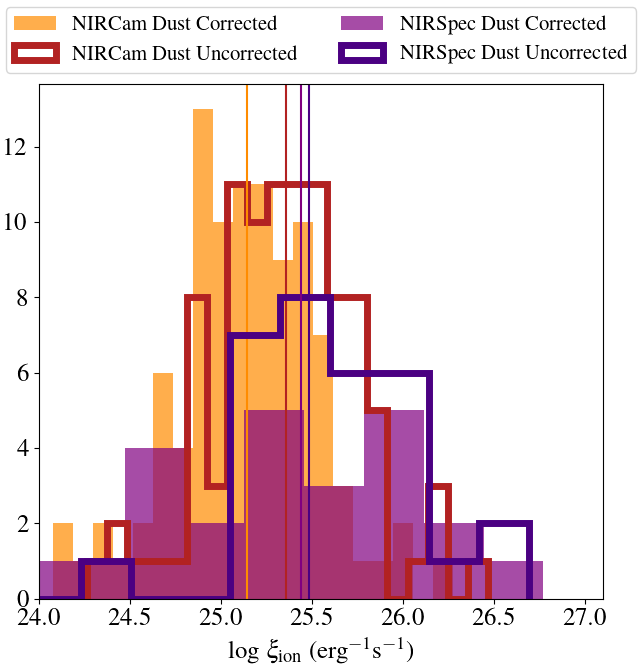}
    \caption{ The distribution of \xiion corrected and uncorrected for dust obscuration is shown in unfilled and filled histograms respectively. The dust correction follows \protect\cite{Calzetti2000} dust law for the NIRCam sample (orange) and using Balmer decrement in the NIRSpec sample (purple). The NIRSpec sample shows no significant difference in the median of the distribution but a slight increase in the $\sigma$ of the distribution. The NIRCam sample shows a lower median \xiion in the dust corrected sample by $\sim 0.2$ dex.} 
    \label{fig:dust}
\end{figure}

\subsection{UV Luminosity and UV continuum slope}
\label{sec:calcuv}
We measure the UV luminosity of galaxies in the NIRCam sample using filters covering rest-frame 1400–1600\AA. Thus we use F090W for galaxies between the redshift range of 4.8 to 5.5 which avoids the Lyman break and Lyman $\alpha$ emission. For the spectroscopic sample,  we use the slit loss corrected 1D spectrum and calculate the UV luminosity in the wavelength window containing 1500 \AA\ determined by the F090W or F115W for each object depending on the redshift. We correct the measured UV luminosity for both photometric and spectroscopic data for lensing magnification. Finally, we bootstrap the measurements within observational errors and report the median values and 1$\sigma$ errors. Figure \ref{fig:spec_phot} compares the UV luminosity measured using photometric and spectroscopic data. We find reasonable agreement between the UV luminosity measured with the two methods with the relative difference being $0.11 \pm 0.45$. 

We calculate \buv from the model SED from EAZY SED fitting of the observed photometry. We measure \buv by fitting model SED between rest-frame 1400–2000 \AA\ with a single power law in the wavelength windows defined by \cite{calzetti1994} to avoid absorption and emission lines. \buv measured with EAZY SED fitting is highly dependent on the templates. The inclusion of the \cite{Larson2022} improves the measurement of \buv by increasing the range to reach up to $\beta_{1500} = -3.1$.

Figure \ref{fig:hist} shows the UV magnitude and \buv distributions before and after magnification corrections. Our sample shows a lower limit (before magnification correction) of \Muv $=-18.1$ (orange unfilled histogram, center) for the NIRCam
sample and \Muv $=-18.45$ (purple unfilled histogram, center) for the NIRSpec sample. After correcting for magnification from lensing from the cluster, the UV magnitude ranges between $-22.68<$ \Muv $<-15.95$ with median \Muv $= -20.07\pm 1.27$ and $-21.7<$ \Muv $<-17.01$ with median \Muv $= -20.1\pm 1.17$ for the NIRCam and NIRSpec samples respectively (filled histograms in \ref{fig:hist}, center ). Given the achromatic nature of lensing, \buv is not affected by lensing magnification. The distribution of \buv shows a median $\beta_{1500} = -2.58 \pm 0.26$ for the NIRCam sample and a median $\beta_{1500} = -2.48 \pm 0.42$ in the NIRSpec sample. After figure \ref{fig:hist}, we only present magnification corrected quantities.

\begin{figure*}
    \includegraphics[scale =0.53]{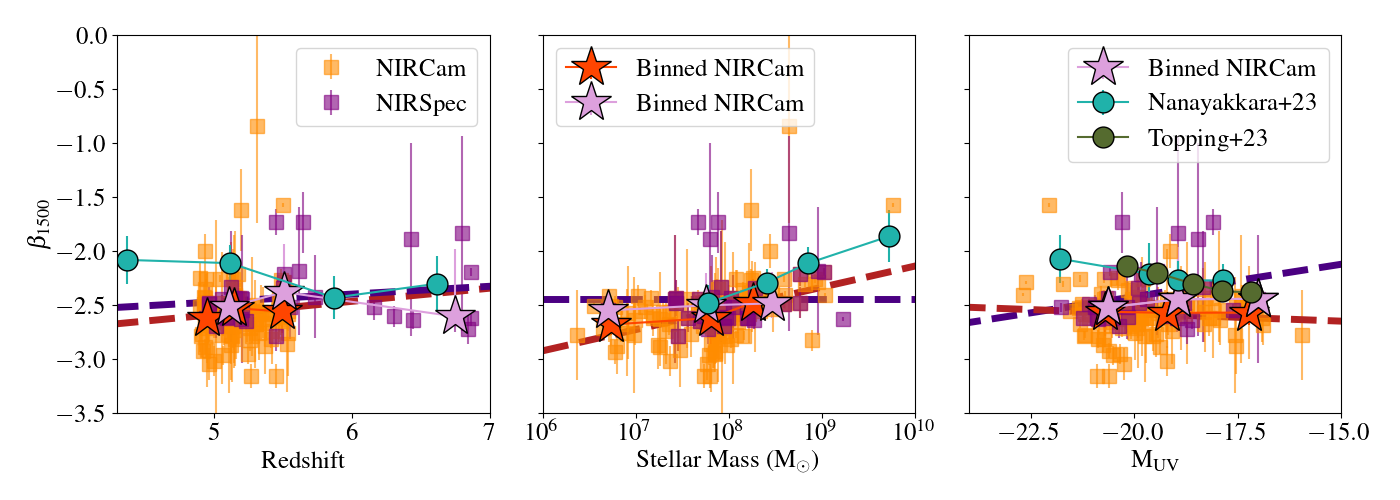}
    \caption{ UV continuum slope (\buv) vs redshift (left), stellar mass (center) and M$_{UV}$ (right). Stars show the binned sample and the dashed lines show the linear fit to the full sample.  Within the uncertainties, we find no significant correlation between \buv and redshift or \Muv. However, we do find a moderate correlation between the \buv and stellar mass in our samples. }
    \label{fig:beta}
\end{figure*}


\subsection{Dust Correction}
The measured H$\alpha$ fluxes are corrected for nebular dust attenuation to account for the light absorbed by dust in the star forming regions in the galaxy. The Balmer decrement (H$\alpha$/H$\beta$) is one of the most robust ways to correct for nebular dust attenuation. To calculate the intrinsic H$\alpha$ luminosity for our spectroscopic sample, we determine the dust attenuation towards H$\alpha$ by calculating the Balmer color excess following the Case B recombination value \citep{Osterbrock1989}. 

\begin{equation*}
    E \left(\rm{H}\beta-\rm{H}\alpha \right) = 2.5 \log\left(\frac{\rm{H}\alpha\rm{/}\rm{H}\beta}{2.86}\right)
\end{equation*}
Following the Calzetti dust law \citep{Calzetti2000}, the intrinsic H$\alpha$ luminosity is determined as: 
\begin{gather*}
    L_{\rm{H}\alpha, int} =  10^{0.4\rm{A}(\rm{H}\alpha)}\times L_{\rm{H}\alpha, obs} \\
    \rm{A}(\rm{H}\alpha) = 6.6 \log\left(\frac{\rm{H}\alpha/\rm{H}\beta}{2.86}\right)
\end{gather*}

For our photometric sample, where we do not have measurements of H$\beta$ flux to determine the balmer decrement, we use the total attenuation ($\rm{A}v$) determined using Dense Basis SED fitting to calculate the attenuation towards H$\alpha$ luminosity. For our sample, we calculate the median $\rm{A}v =  0.11^{+0.04}_{-0.06}$. We adopt the Calzetti dust law following the recent results from \cite{markov2024}, who find the dust law for galaxies at $\rm{z}>4.5$ is comparable to the Calzetti dust law. Following Calzetti dust law \citep{Calzetti2000}, attenuation towards H$\alpha$ is determined as: 
\begin{gather*}
    \rm{A}(\rm{H}\alpha) = 0.82\times \rm{A}v
\end{gather*}

Given the dust attenuation is higher in the UV wavelengths, the observed UV luminosity in both photometric and spectroscopic data is similarly corrected to determine the intrinsic UV luminosity using the Calzetti dust law at $\lambda = 1500$\AA\, as:
\begin{gather*}
    \rm{A}(1500) = 2.55\times \rm{A}v
\end{gather*}




\subsection{\xiion Measurement}
\label{sec:xiion}
Lyman-continuum photon production eﬃciency in the presence of a non-zero $f_{esc,LyC}$ is used to determine the contribution of an object in ionising the IGM. Since the precise value of $f_{esc,LyC}$ is uncertain, we assume a $f_{esc,LyC} = 0$. From indirect measurements of $f_{esc}$, studies find median predicted $f_{esc,LyC)}<0.2$  at $6<\rm{z}<9$ \citep{Mascia2023,Saxena2023}. 


\cite{Leitherer1995} found the following relation between the H$\alpha$ luminosity and the intrinsic Lyman-continuum photons production rate computed for dust free Case B recombination \citep{Osterbrock1989} assuming no escape of ionizing photons:

\begin{equation*}
    \rm{L}(\rm{H}\alpha) [erg s^{-1}] = 1.36 \times 10^{-12} \dot{N}_{(H_0)}[s^{-1}].
\end{equation*}
Thus, we express \xiion$_0$ as the ratio of the Lyman-continuum photons production rate to the intrinsic rest-frame UV luminosity ($\rm{L}_{UV}$) as: 
\begin{equation*}
\xi_{\rm{ion,0}} = \frac{\dot{N}_{(H_0)}}{\rm{L}_{UV}}
\end{equation*}

We show the distribution of dust corrected and uncorrected  $\log$ \xiion for the NIRCam and NIRSpec samples in figure \ref{fig:dust}. We find that for our NIRCam sample on average, dust corrected \xiion is $0.20$ dex lower than the uncorrected \xiion. We note that for this sample we do not have information on the Balmer decrement. In our NIRSpec sample, for which we have the Balmer decrement, the dust corrected \xiion is $<0.1$ dex higher than the uncorrected \xiion. Similarly, if we assume SMC-like dust attenuation, dust corrected \xiion is $0.2$  and 0.1 dex lower in the NIRCam and NIRSpec samples respectively. Due to the lack of significant effect of dust on the median \xiion, we report the dust uncorrected \xiion in the paper. 

\section{Results}
\subsection{The UV slope}
\label{sec:uvslope}
The UV continuum slope (\buv) of a galaxy traces the light emitted by young massive stars thus is crucial to constrain the stellar population and star formation history of a galaxy. \buv is dependent on the ages, metallicities, and dust content in a galaxy \citep{ Duncan2015, Reddy2018}. Given its dependence on the dust content of the galaxy,  \buv is often used to constrain the dust attenuation of the spectrum where Balmer decrement or infrared spectrum is not observed. However, intrinsically, for a dust corrected spectrum, the \buv is dependent on the fraction of very young and metal-poor stellar population. Specifically, young and metal-poor stellar populations result in bluer \buv compared to older metal-rich stellar populations. Given that the main source of ionising UV radiation in a star forming galaxy is from the young massive stellar population, the production efficiency of ionising UV radiation (\xiion) is expected to correlate with the \buv. In this section, we study the distribution and the evolution of \buv with galaxy properties.

In section \ref{sec:calcuv} we describe the method we have adopted to measure \buv for the NIRCam samples. In figure \ref{fig:hist}, we present the distribution of \buv with 156 galaxies in the photometric (orange). We find   $-3.15<\beta_{1500} <-0.83$  with the median at $\beta_{1500} = -2.56 \pm 0.26$ in the photometric sample. In figure \ref{fig:beta}, we plot \buv as a function of redshift, stellar mass, and \Muv for both the NIRCam (orange) and NIRSpec (purple) samples. The binned sample along with the linear regression between \buv and redshift, stellar mass, and \Muv are also shown. 

\textbf{Redshift Evolution of \buv:} We find no significant evolution of \buv with the redshift in the NIRCam sample (slope: 0.12, Spearman Coefficient: 0.10). However, we note that our NIRCam sample is restricted to a short range of redshift between $4.8<\rm{z}<5.5$. In our NIRSpec sample, the linear regression measures a slope of 0.07 with a large scatter. Thus  we do not find a significant correlation between \buv and z. This overall trend is comparable to the results in the literature \citep{Topping2023,Nanayakkara2023,Jiang2020,Bouwens2014}. We note a large scatter in our sample and a lower correlation compared to the empirical relation of \buv with redshift shown by \cite{Bouwens2014}. We also note that previous studies like \cite{Jiang2020} are unable to recover $\beta_{1500}<-2.6$ with SED fitting due to the limitations of the then available templates, whereas \buv calculated using broad band photometry at high redshift \citep{Bouwens2014} are affected by the uncertainties in redshifts as well as spectral features. 

\textbf{Correlation of \buv with Stellar Mass:} In the middle panel of figure \ref{fig:beta}, we present the correlation of \buv with the stellar mass of the galaxy. The stellar mass (magnification corrected) in our NIRCam sample ranges from \Mstar = 6.3 to 9.75. We find a moderate correlation of \buv with the magnification corrected stellar mass of the galaxies in both our NIRCam sample such that \buv gets shallower with increasing stellar mass. The linear regression of \buv with stellar mass for the NIRCam sample gives a slope of 0.19 and a Spearman coefficient of 0.46, and a slope of 0.007 with low significance in the NIRSpec sample. In the NIRCam sample, the lower mass sample of \Mstar$<8$ (median = $7.76\pm 0.04$) has a \buv $= -2.62\pm 0.02$ compared to \buv $= -2.46\pm 0.04$ in the higher mass sample \Mstar$\geq8$ (median = $8.26\pm 0.05$). The bluer \buv of low mass galaxies implies that the low mass galaxies could have lower dust content, possible higher fraction of ionising radiation, or lower metallicities compared to high mass galaxies. Bluer \buv is also associated with high escape fraction \citep{Chisholm2022, Mascia2023}, indicating a higher escape fraction in low mass galaxies and thus a higher contribution of low stellar mass galaxies during the epoch of reionisation. This result is comparable to \cite{Nanayakkara2023} who found a correlation between \buv and the stellar mass of galaxies at $4<\rm{z}<8$ similar to our NIRCam sample. We also note that the \buv calculated in \cite{Nanayakkara2023} follows a similar procedure to measure \buv as our NIRCam sample but they do not perform magnification correction for the stellar mass. 

\textbf{Correlation of \buv with UV Magnitude:} Finally, we present the correlation of UV magnitude to the \buv in the right panel in figure \ref{fig:beta}. For our  sample with UV magnitude $\rm{M}_{UV}<-19$, we find a median \buv $=-2.58\pm0.02$ and for the fainter sample $\rm{M}_{UV}>-19$, we median find \buv $= -2.54\pm0.02$. In our NIRCam sample, we do not observe a significant UV magnitude evolution of the \buv (with slope $= -0.01$ and the Spearman coefficient= -0.02).  Similarly, in the NIRSpec sample, we do not observe a significant UV magnitude evolution of the \buv with slope $= 0.05$ and Spearman coefficient = 0.03.

We find correlation presented in \cite{Nanayakkara2023} and \cite{Bouwens2014} for a sample at z $=4$ to 7, and \cite{Topping2023} for a sample at z $= 5$ to 7 comparable to our NIRCam sample, albeit with a shallower slope of the linear fit. The binned median \buv in \cite{Nanayakkara2023} from the GLASS JWST survey at a given UV magnitude is higher compared to our NIRCam sample. However, this discrepancy could be explained by the photometric redshift selection of galaxies in \cite{Nanayakkara2023}, whereas we have an additional condition of F410M flux excess from H$\alpha$ emission which biases our sample towards actively star forming galaxies. Similarly, \cite{Topping2023}, who calculate \buv  directly from photometry also have higher median \buv values at UV magnitude $\rm{M}_{UV}<-19$ which could be driven by the Lyman break selection. However, previous studies like \cite{yamanaka2019} find an opposite correlation between \buv and $\rm{M}_{UV}$ in the range ($-22>\rm{M}_{UV}>-20$) at z $\sim 4$. The Lyman break selected galaxies in \cite{yamanaka2019} are possibly biased towards higher dust content. 

\subsection{Ionising Photon Production Efficiency \xiion}
\begin{figure*}
    \includegraphics[scale = 0.55]{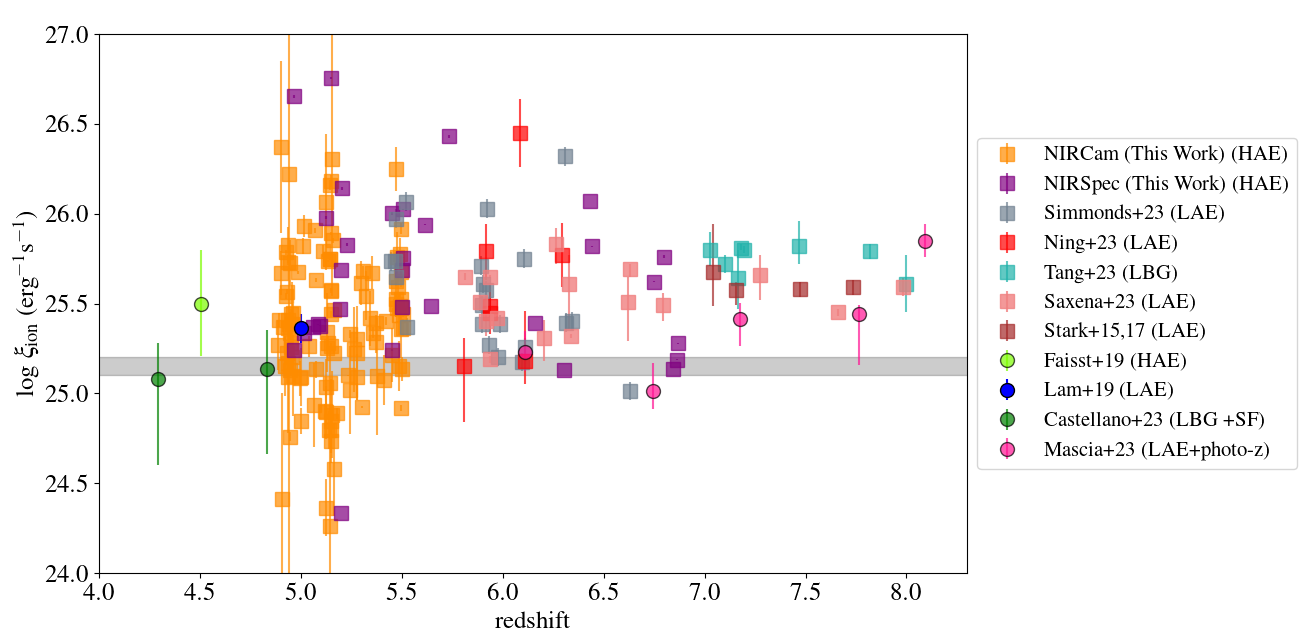}
    \caption{Evolution of the ionising photon production efficiency \xiion (uncorrected for dust) with redshift in the epoch of reionisation. The purple and orange squares of NIRSpec and NIRCam selected samples respectively are shown with various measurements of \xiion from the literature. The method of target selection is mentioned in parentheses. Circles show binned measurements of $\log$ \xiion, whereas squares show individual galaxy measurements. The grey line shows the linear fit to the canonical value of \xiion required to ionise the local IGM as calculated by \protect\cite{Robertson2015}. There is no significant evolution of \xiion during the epoch of reionisation and the \xiion of our sample at $\rm{z}<5.5$ is comparable to their higher redshift counterparts from the literature within the errors.}
    \label{fig:xion_z}
\end{figure*}

Constraining \xiion during the epoch of reionisation for different galaxy populations is important to understand the contribution of different galaxies to EoR and to understand how the reionisation of the Universe proceeded. In this section we explore the evolution of \xiion with redshift and other galaxy properties at z $\sim$ 4.8 to 7 in our NIRCam and NIRSpec samples.

\subsubsection{Evolution of \xiion}
We calculate the median \xiion of H$\alpha$ selected NIRCam sample in the redshift range $4.8<\rm{z}<5.5$ $\log\,$\xiion $=25.36\pm0.40$ compared to the NIRSpec sample in the same redshift range with $\log\,$\xiion $=25.42\pm0.3$. In this redshift range, we find comparable results (within $1\sigma$) between our NIRCam and NIRSpec samples. Whereas, the median \xiion for galaxies in the NIRSpec sample in redshift range $5.5<\rm{z}<7$ is $\log\,$\xiion $=25.5\pm0.2$ which is higher than the median \xiion at $4.8<\rm{z}<5.5$ by $<1\sigma$.


We put in context our measurements by comparing with H$\alpha$ emitters at $4<\rm{z}<4.5$ from \cite{Faisst2019}, Lyman-break galaxies at $4<\rm{z}<6$ from \cite{Bouwens2016}, [OIII] emitters at $5<\rm{z}<9$ from \cite{Matthee2023}, \cite{Tang2023} and \cite{Mascia2023}(also H$\alpha$ and Ly $\alpha$ emitters), and Lyman-alpha emitters (LAEs) at $\rm{z}>4$ from \cite{Lam2019}, \cite{Ning2023}, \cite{simmonds2023} and \cite{Saxena2023}. Among these,\cite{Matthee2023}, \cite{Tang2023} and \cite{Saxena2023} measure Balmer line fluxes directly from spectroscopy and \cite{Bouwens2016}, \cite{Faisst2019}, \cite{Lam2019}, \cite{Ning2023} and \cite{simmonds2023} make measurements from photometry.

The median $\log\,$\xiion $=25.36\pm0.04$ at $4.8<\rm{z}<5.5$ in our photometric sample is comparable to the median $\log$ \xiion $=25.27\pm0.03, 25.5, 25.36\pm0.08 $ reported in \cite{Bouwens2016}, \cite{Faisst2019} and \cite{Lam2019} at similar redshifts. At $\rm{z}\sim6$, \xiion from \cite{Ning2023} and \cite{simmonds2023} (median $\log$ \xiion $=25.48\pm0.4$ and $25.44^{+0.21}_{-0.15}$ respectively) are also comparable to our \xiion at $\rm{z}<5.5$. At $5.5<\rm{z}<7$, the median $\log$ \xiion of $=25.6\pm0.2$ in our NIRSpec sample is comparable to $\log\,$\xiion $=25.3, 25.8 \pm0.09$ and $25.56$ at  $6<\rm{z}<8$ reported in \cite{Matthee2023},\cite{Tang2023} and \cite{Saxena2023}. 

From our NIRSpec sample, we confirm that 4 out of 33 galaxies are Ly $\alpha$ emitters (LAE) with median $\log$ \xiion = $25.49\pm0.17$ compared to the median $\log\,$\xiion $25.48=\pm0.2$ of the non-LAE. We find no significant difference between the \xiion of LAE and non-LAE in our NIRSpec sample. We also find no significant difference between LAE reported in \cite{Lam2019}, \cite{Ning2023}, \cite{simmonds2023}, and \cite{Saxena2023} and the non-LAE in our sample. The lack of significant difference between \xiion in the LAE and non-LAE also indicates different physical processes driving the production of ionising photons and creating channels of escape for Lyman $\alpha$ photons in different galaxies similar to the findings in \cite{Saxena2023}. 

Overall, we find marginal evolution of median \xiion between our sample and median \xiion reported in the literature between $4<\rm{z}<9$. At $\rm{z}<5.5$. Median $\log\,$\xiion $=25.36\pm0.04$ and $\log\,$\xiion $=25.48\pm1$ for our NIRCam and NIRSpec sample respectively is lower ($< 1 \sigma$) than $25.8 \pm0.09$ and $25.56$ reported in \cite{Tang2023} and \cite{Saxena2023}. However, it is consistent with $\log\,$\xiion $=25.36\pm0.07$ and 25.3 in \citep{Mascia2023} and \cite{Matthee2023} respectively. We note that the selection of H$\alpha$ emitters for our sample biases the sample towards higher \xiion compared to a LBG selection that is done in the stated literature studies and we might miss galaxies with low \xiion. We also find a large scatter in \xiion, especially in the NIRCam sample, which indicates a stochastic star formation in galaxies at $z>5$.

\subsubsection{\xiion vs Stellar Mass and Star Formation}

\begin{figure*}
    \includegraphics[width = \textwidth]{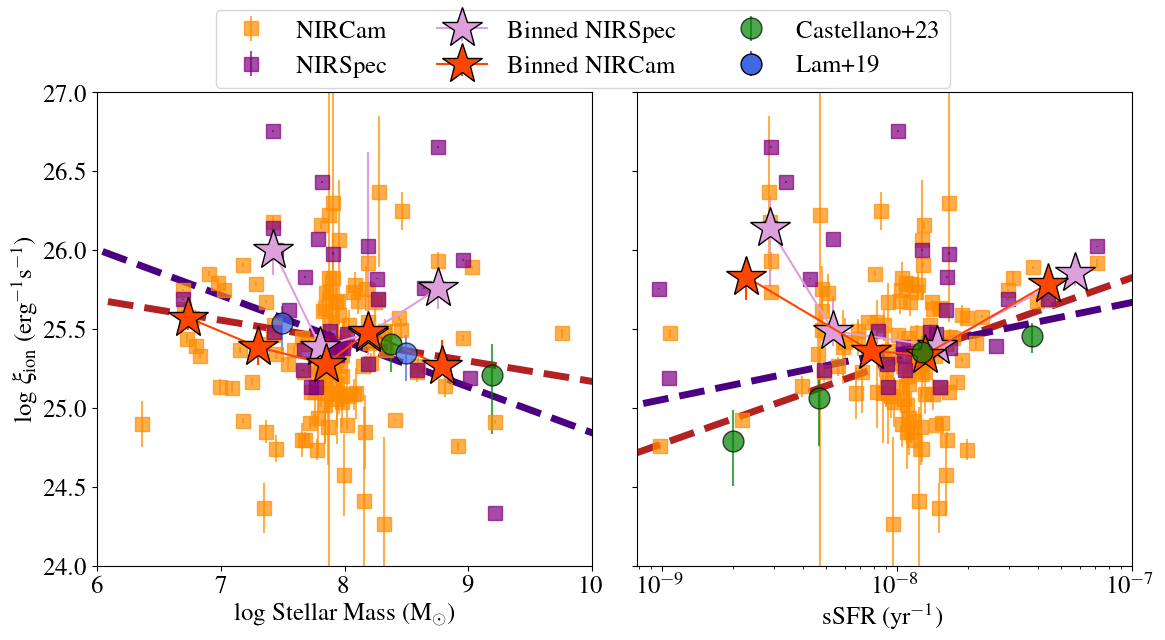}
    \caption{The ionising photon production efficiency (\xiion) as a function of stellar mass (left) and the specific star formation rate (right). The purple and orange colors show our NIRSpec and NIRCam samples respectively. The squares and stars show individual galaxies and binned samples with the lines showing linear fits to the sample. We find a moderate negative correlation of \xiion with the stellar mass of galaxies and a moderate positive correlation of \xiion with the sSFR. Dwarf star forming galaxies have a higher photon production efficiency and will require a lower escape fraction to ionise their local IGM.} 
    \label{fig:sm}
\end{figure*}
\begin{figure*}
    \includegraphics[width = 0.5\textwidth]{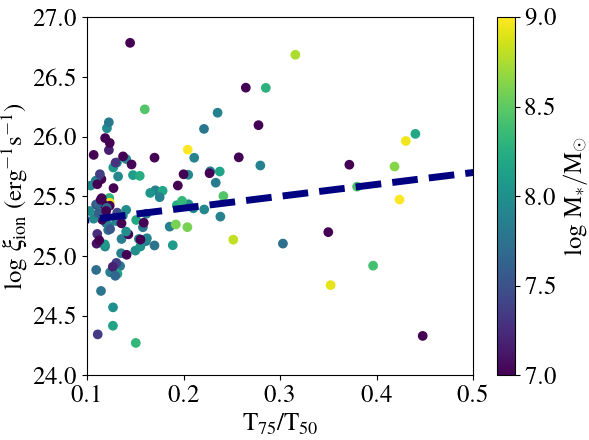}
    \includegraphics[width = 0.48\textwidth]{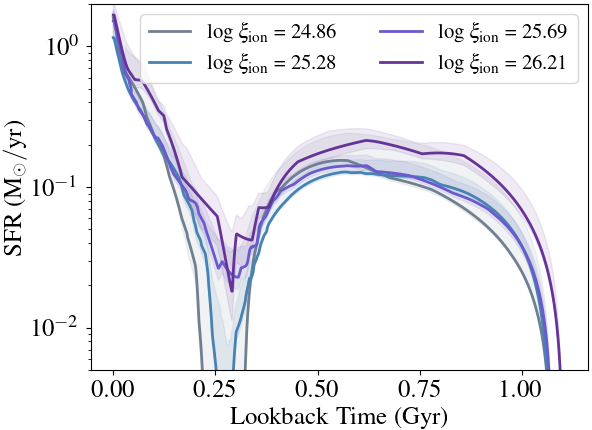}
    \caption{The ionising photon production efficiency (\xiion) is slightly positively correlated with T$_{75}/$T$_{50}$ (left). A more recent burst in star formation has a smaller T$_{75}/$T$_{50}$. The markers have been  color-coded with the magnification-corrected stellar mass of the galaxy. The right panel shows the star formation history of galaxies in bins of \xiion with median \xiion of the bin stated in the legend.  } 
    \label{fig:sfh}
\end{figure*}

\begin{figure*}[!h]
    \includegraphics[width = \textwidth]{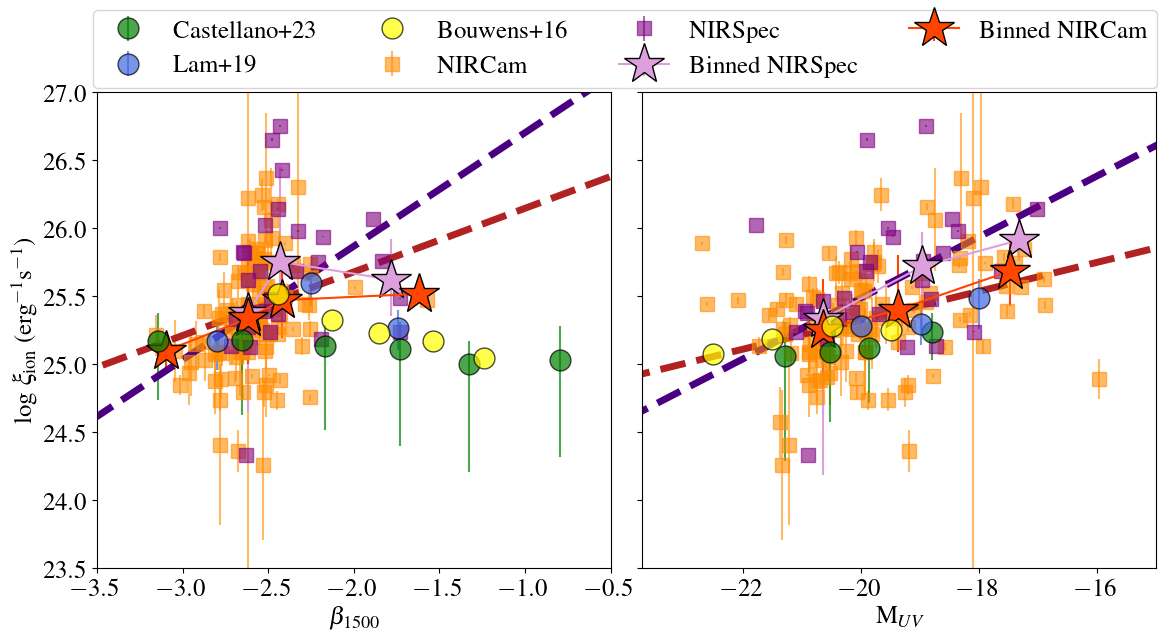}
    
    \caption{The ionising photon production efficiency (\xiion) as a function of UV slope (\buv) (left) and the UV magnitude ($\rm{M}_{UV}$)  (right). The purple and orange colors show the NIRSpec and NIRCam samples respectively. The squares and stars show individual galaxies and binned samples with the lines showing linear fits to the sample. We find a positive correlation of \xiion with the $\rm{M}_{UV}$ in both NIRCam and NIRSpec samples similar to past studies. However, we find no significant correlation between \xiion and \buv. } 
    \label{fig:uv}
\end{figure*}

The dependence of the production of ionising photons on the young and massive stellar population in a galaxy indicates that \xiion should show a correlation with the stellar mass and the star formation history of a galaxy. In this section, we explore the evolution of \xiion with the stellar mass and star formation history of the galaxies.

Figure \ref{fig:sm} shows \xiion as a function of magnification-corrected stellar mass and the specific star formation rates calculated from Dense Basis SED fitting of the NIRCam (orange squares) and NIRSpec sample (purple squares). The NIRCam sample covers a range of magnification corrected \Mstar $=6.3 - 9.8$ and the NIRSpec sample covers the range of \Mstar $= 7.2 - 9.1$. We further bin the sample into 5 stellar mass bins of \Mstar $=6.3 - 7$,  $7 - 7.5$,  $7.5- 8$, $8- 8.5$ and $>8.5$. The stars show the median \xiion and the median stellar mass of the binned sample. We find that \xiion decreases marginally (by $\sim 0.12$ dex) with increasing stellar mass such that the median $\log\,$\xiion $= 25.57\pm0.06$ for galaxies with \Mstar $<7$ and $=  25.46\pm0.1$ for \Mstar $>8$. The linear fit also shows a negative slope of -0.16 and -0.28 for the NIRCam and NIRSpec samples respectively. Although the negative slopes indicate a moderate negative correlation between \xiion and the stellar mass of the galaxy, we note that the intrinsic scatter of the \xiion measurement is large at standard deviation $1\sigma = 0.40$ for the NIRCam sample and $1\sigma = 1.03$ for the NIRSpec sample. Further, we calculate the Spearman correlation coefficient. We find a negative correlation coefficient of -0.03 for the NIRCam sample and -0.21 for the NIRSpec sample. Our NIRSpec sample indicates a slight negative correlation of \xiion with the stellar mass of the galaxies within a large intrinsic scatter, however, the Spearman correlation finds no significant correlation between \xiion and stellar mass in our NIRCam sample. 

\begin{figure*}
    \includegraphics[width = \textwidth]{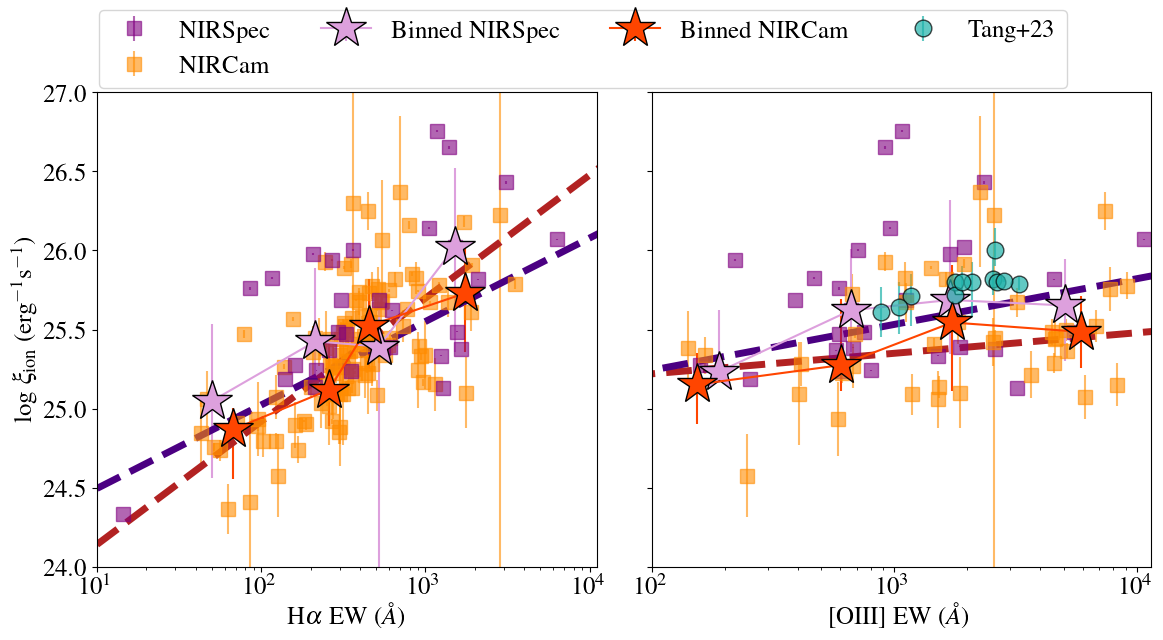}
    \caption{ The ionising photon production efficiency (\xiion) as a function of H$\alpha$ equivalent width (left) and  [OIII] equivalent width  (right). The purple and orange colors show our NIRSpec and NIRCam samples respectively. The [OIII] equivalent width is only shown for the galaxies in the NCF field which have medium band observation of 5007 \AA\ restframe wavelength. The squares and stars show individual galaxies and binned samples with the lines showing linear fits to the sample. We find a positive correlation of \xiion with the H$\alpha$ equivalent width. We also find a moderate positive correlation of \xiion with the [OIII] equivalent width.}
    \label{fig:ew}
\end{figure*}
Given the difficulty in observing faint low stellar mass galaxies at $z\gtrsim5$, which has only recently been made available with the launch of JWST, similar results have been reported in the literature albeit at lower redshifts. \cite{Lam2019} and \cite{Castellano2022} find similar slopes of the linear fit for the correlation between \xiion and stellar mass at \Mstar $>7$ and \Mstar $>8$ at redshift $\sim 5$ respectively. Compared to this sample, we can extend the range of stellar mass by $>1$ dex towards the lower mass end. The marginal elevation \xiion of the lower mass galaxies has important implications for the progression of ionisation and the sources of ionising photons during the epoch of reionisation. The steep slope of the faint end of the UV luminosity function at $z>5$ \citep{Bouwens2015, Livermore2017, Atek2018, Atek2023} coupled with higher \xiion of low mass galaxies indicate such dwarf systems may emit significant ionising photons into the IGM to reionise their local environment depending on their escape fraction of ionising photons. However, we note that our sample which is biased towards high \xiion galaxies especially at the magnification uncorrected low stellar mass end due to the H$\alpha$ selection could be driving this difference.  

In the right panel of Figure \ref{fig:sm}, we show \xiion as a function of the specific star formation rates (sSFR) of the galaxies as measured from SED fitting (refer to section \ref{sec:sedfit} for the description of SED fitting). We note that both \xiion and sSFR are not affected by magnification correction. Our samples cover a range of log sSFR =  -9 to -7 $\rm{yr}^{-1}$. Similar to the stellar mass, we have binned the sample in sSFR and the stars depict the median \xiion vs the median sSFR of the binned sample. We bin the sample with log sSFR = $-9\, \rm{to} -8.5$,  $-8.5\, \rm{to} -8$,  $-8\, \rm{to} -7.5$ and $>-7.5$ $\rm{yr}^{-1}$. In both the NIRCam and NIRSpec samples, we find a positive correlation between \xiion and sSFR with slope of the linear fit 0.48 for the NIRCam sample and 0.29 for the NIRSpec sample. However, we also find a small sub-sample of outliers to this correlation at the low sSFR end (log sSFR $<8.5 \rm{yr}^{-1}$). This low sSFR end of our sample ($\sim10 \%$ of the sample) shows a larger scatter and higher median \xiion. This difference could be a result of the H$\alpha$ selection, which would bias our sample to higher \xiion, especially at low sSFR. We note that between log sSFR -8.5 to -7.5, where $\sim85 \%$ of the sample lies, we find no significant correlation of \xiion but a large intrinsic scatter of $\sigma = 0.38$.  We further calculate the Spearman coefficient of -0.06 and -0.13 for correlation between \xiion and log sSFR for the NIRCam sample and the NIRSpec sample respectively. Although the linear fit and the Spearman coefficient point towards a slight positive correlation between \xiion and the sSFR of a galaxy, we note that these results have high intrinsic scatter and do not show a clear correlation.

We compare our results with \cite{Castellano2023}, who find a very significant monotonically increasing correlation between \xiion and sSFR of galaxies at $2<\rm{z}<5$. They report that the \xiion increases from  $\log$ \xiion $= 24.5$ \xiunit at $\log \rm{sSFR} = -9.5\, \rm{yr}^{-1}$ to $= 25.5$ \xiunit at $\log \rm{sSFR} = -7.5 \,\rm{yr}^{-1}$. They also found a Spearman correlation coefficient of $\sim 0.79$ compared to 0.013 and a large p-value of 0.5 in our sample. Similar to our results, \cite{Castellano2023} also find a larger scatter of \xiion at the lower end of sSFR indicating a lower correlation of \xiion with sSFR at the low sSFR end, however, unlike our result, the median \xiion at the lower sSFR is lower than \xiion at the higher sSFR end. The difference in the results could be driven by the H$\alpha$ selection or could also be explained by the redshift evolution of the median sSFR of the galaxies indicating that at $\rm{z}>5$, \xiion is not highly correlated to the sSFR of the galaxy as galaxies may have a stochastic star formation history which will rapidly fluctuate the sSFR of a galaxy. 

Given the possible effect of stochasticity, we explore further the time scales with which \xiion changes with the star formation by studying the correlation of \xiion with stellar ages and star formation histories. We derive the times when the galaxy creates 50\% and 75\% of its stellar mass ($\rm{T}_{50}$ and $\rm{T}_{75}$, at the time of observation) using the non-parametric histories of Dense Basis SED fitting. Figure \ref{fig:sfh} shows \xiion as a function of the ratio  $\rm{T}_{75}$/$\rm{T}_{50}$ (left panel) with each galaxy colored by its stellar mass. The ratio $\rm{T}_{75}$/$\rm{T}_{50}$ is indicative of the relative time of the last burst of star formation in the galaxy such that a lower $\rm{T}_{75}$/$\rm{T}_{50}$ indicates a more recent burst. We find a moderate positive correlation between \xiion and  $\rm{T}_{50}$/$\rm{T}_{75} <0.3$ such that \xiion is decreasing on average with more recent bursts. However, we find a small sample of galaxies with $\rm{T}_{50}$/$\rm{T}_{75}>0.3$ that have a lower median \xiion.  This indicates a potential correlation of \xiion with the age of the starburst, however, we note that the effect is also dependent on the star formation history and the strength of the starburst. We also find the lowest stellar mass galaxies have had a more recent burst compared to the more massive galaxies in our sample. We note that the H$\alpha$ selection biases our sample to galaxies with ongoing active star formation i.e. either to a rising or bursty star formation history. 

We further divide our sample into \xiion bins and plot their median star formation histories (right panel, figure \ref{fig:sfh}). The median \xiion of each bin and their respective medians magnification corrected stellar masses are: $\log$ \xiion = $24.86\pm0.05,\, 25.28\pm0.02,\, 25.69 \pm0.02,\, 26.21\pm0.04$ and \Mstar = $7.78\pm0.14,\, 7.89\pm0.07, 7.88\pm0.1,\, 7.90 \pm 0.12$. We find that on average our sample is experiencing an ongoing star formation burst. However, this could be a result of the selection of H$\alpha$ emitters exclusively. The bin with the highest \xiion started the burst $<100$ Myrs earlier than the lowest \xiion bin. Thus, we infer that there may be a potential delay of increase in the production rate of ionising photons from the start of the last rise in star formation.

 
\begin{figure*}
    \includegraphics[width = \textwidth]{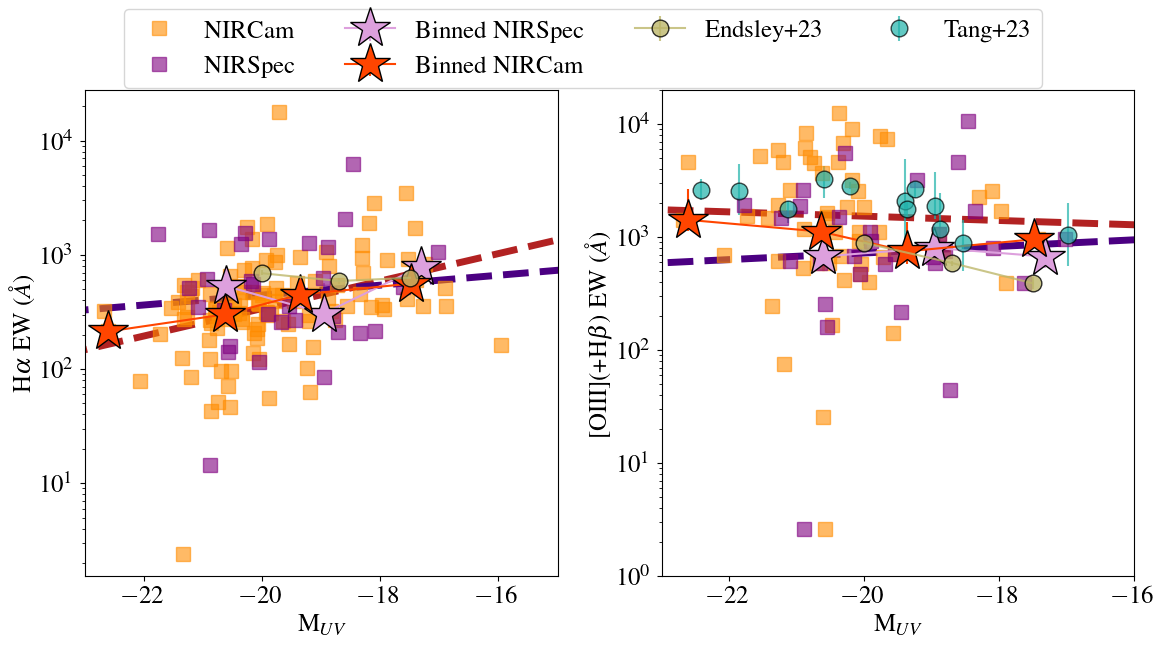}

    \caption{ H$\alpha$ equivalent width (left) and [OIII](+H$\beta$) equivalent widths (right) as a function of UV magnitude. The purple and orange colors show our NIRSpec and NIRCam samples respectively. The [OIII] equivalent width is shown for the NIRSpec sample and [OIII]+H$\beta$ EW for galaxies in the NCF field which have medium band observation of 5007 \AA\ restframe wavelength. The squares and stars show individual galaxies and binned samples with the lines showing linear fits to the sample. We find a positive correlation of  H$\alpha$ EW with the UV magnitude and no significant correlation of [OIII](+H$\beta$) EW with the UV magnitude.} 
    \label{fig:ewmuv}
\end{figure*}

\subsubsection{\xiion vs UV properties}

In this section, we study the correlation of UV properties with \xiion. In figure \ref{fig:uv}, we show \xiion as a function of the UV slope (\buv, left panel) and the UV magnitude ($\rm{M}_{\rm{UV}}$, right panel). We find that \xiion correlates positively with \buv in our NIRCam sample with slope = 0.49 and the Spearman correlation coefficient = 0.32. Similarly, \xiion correlates positively with \buv with slope of 0.5 in out NIRSpec sample. However, the \buv measurement from EAZY SED fitting recovers \buv in a small range ($-3.1<$\buv$<2$), and we find a significant scatter of \xiion within this range. Compared to other studies, \cite{Bouwens2016} and \cite{Castellano2023} find a negative correlation of \buv with \xiion , whereas \cite{Lam2019} find no significant correlation. From the variability of \buv with measurement and the difference in results across literature, we infer that \xiion is not correlated with \buv \, at $\rm{z}>5$. 

We measure the correlation between \xiion and the $\rm{M}_{\rm{UV}}$. We find a positive correlation with a slope and Spearman correlation coefficients of 0.16 and 0.34 in our NIRCam sample, and 0.18 and 0.61 in our NIRspec sample. We find a steeper correlation between \xiion and $\rm{M}_{\rm{UV}}$ compared to \cite{Bouwens2016}, \cite{Castellano2023} and \cite{Lam2019}. We are able to increase the range of $\rm{M}_{\rm{UV}}$ towards the faint end to $\sim -16$ and our result indicates a higher production rate of ionising photons for UV faint galaxies compared to other studies.

\subsection{Ionisation Conditions from restframe Emission Lines}

The ionising photon production efficiency depends on the star formation and the ionising conditions of the galaxy ISM. In this section, we study the correlation between \xiion and H$\alpha$ and [OIII](+H$\beta$) equivalent widths which trace the average stellar ages and the ionisation parameter of the ISM respectively. Figure \ref{fig:ew} shows \xiion as a function of H$\alpha$ EW (left) and [OIII](+H$\beta$) EW (right). \xiion is highly correlated with H$\alpha$ EW in both NIRCam (slope 0.7, Spearman coefficient 0.6) and NIRSpec (slope 0.5, Spearman coefficient 0.5) samples. Although partially the correlation found is driven by the fact that \xiion is measured with H$\alpha$ luminosity, \xiion is also dependent on UV luminosity, and H$\alpha$ EW is dependent on the strength of the stellar continuum as well. Our results are comparable with \cite{Atek2022}, who find a correlation with a slope of 0.63 for low stellar mass galaxies at z $\sim1$ and other studies at low redshifts ( z $\sim1-2$) \citep{Emami2020,
Tang2019,Matthee2017}. 

At lower redshifts (z $\sim1-2$), \xiion is known to correlate with the  [OIII]+H$\beta$ EW \citep{Chevallard2018,Tang2019,Onodera2020}. At higher redshifts \cite{Tang2023} show for a small sample of $z>7$ galaxies that the \xiion is similarly correlated with [OIII]+H$\beta$ EW and also that [OIII]+H$\beta$ EW is correlated with the O32 ratio which traces the ionisation parameter of the galaxy ISM. Here we investigate the correlation of \xiion [OIII](+H$\beta$) EW in our NIRSpec and NIRCam (only in the NCF field) samples. We find that, \xiion is moderately correlated with [OIII]+H$\beta$ EW in the NIRCam sample  (slope 0.12, Spearman coefficient 0.5) and with [OIII] EW in the NIRSpec sample (slope 0.4, Spearman coefficient 0.3). This result is consistent with the correlation found in both lower and higher redshift samples \citep{Chevallard2018,Tang2019,Onodera2020, Tang2023}. In conjunction with the result in \cite{Tang2023}, this indicates that \xiion is correlated with the ionisation parameter of the galaxy.

In the NIRCam sample, we find median [OIII]+H$\beta$ EW $=1861\pm433$ \AA\  which is comparable to $1993\pm214$ \AA\ for [OIII] emitters at z $\sim8$ \citep{Tang2023} but higher than the $\sim 700$ \AA\ at z $\sim 6$ found in \cite{Endsley2023}. The higher [OIII]+H$\beta$ EW in our sample compared to \cite{Endsley2023} could be driven by our H$\alpha$ selection which biases the sample towards higher star forming and hence emission line galaxies. In our NIRSpec sample, we find a median [OIII] EW of $802\pm360$\AA\ which is higher compared to the lower redshift sample ($\sim$100 \AA\ for star forming galaxies at z$=2$ \citep{Reddy2018}), but similar to the [OIII] EWs in extreme emission line galaxies at z $\sim3$ \citep{Gupta2023}.

Further, we investigate the correlation of H$\alpha$ and [OIII]+H$\beta$ EWs with the UV magnitude shown in figure \ref{fig:ewmuv}. We find a positive correlation between H$\alpha$ and the M$_{\rm{UV}}$ in the NIRCam sample with a slope of 0.12 and a Spearman correlation coefficient of 0.41 with a very small p-value indicating a strong correlation. However, we find no significant correlation between H$\alpha$ and the M$_{\rm{UV}}$ in the NIRSpec sample with 0 slope and a large p-value for the Spearman correlation coefficient. The difference in correlation of H$\alpha$ with M$_{\rm{UV}}$ between the NIRSpec and NIRCam sample might be driven by the H$\alpha$ selection and shallower depth of the NIRSpec sample compared to the NIRCam sample. We note that at M$_{\rm{UV}} <-18$, we also find a large scatter in H$\alpha$ (median H$\alpha$ EW $= 339.27$\AA\, and $\sigma = 1651.06$\AA) compared to the fainter end (M$_{\rm{UV}} >-18$; median H$\alpha$ EW $= 517.8$\AA\, and $\sigma = 794.69$\AA). Although the UV faint sample (M$_{\rm{UV}} >-18$) has been magnified due to gravitational lensing from the foreground cluster, H$\alpha$ EW is independent of the magnification. Our results are comparable to the median H$\alpha$ EW ($\approx 400$ \AA) found in previous studies \citep{Faisst2019,Lam2019}. However, we report a lower median H$\alpha$ EW compared to EW reported in \cite{Endsley2023a}, who also do not find a correlation between H$\alpha$ EW and M$_{\rm{UV}}$ in the Lyman break selected galaxies.

Figure \ref{fig:ewmuv} shows the correlation between [OIII]+H$\beta$ EW and the UV magnitude. We find no significant correlation between [OIII]+H$\beta$ EW in the bright UV sample (M$_{\rm{UV}} <-18$). Due to the lack of medium bands observations of the [OIII]+H$\beta$ emission in the CLU field which are magnified due to gravitational lensing, we are limited to the M$_{\rm{UV}} <-18$ sample. Within the range $-23< $M$_{\rm{UV}} <-18$, we find a median [OIII]+H$\beta$ EW $=917.27$ \AA\ and a large scatter of $\sigma = 2621.65$\AA. We also do not find a correlation between [OIII]+H$\beta$ EW  and the UV magnitude with 0 slope and large p-value for Spearman correlation coefficients. Similar to previous studies, we find a large [OIII]+H$\beta$ EW compared to z$\sim2$, and 46 extreme emission galaxies with $>1200$ \AA\, [OIII]+H$\beta$ EW. The median  [OIII]+H$\beta$ EW of our NIRCam sample is lower than 1994 \AA\, reported in \cite{Tang2023} for z$>7$ galaxies but higher than median [OIII]+H$\beta$ EW ($<890\pm60$ \AA) reported in \cite{Endsley2023a}. \cite{Endsley2023a} also find that the UV faint (M$_{\rm{UV}} >-19$) galaxies have a lower [OIII]+H$\beta$ EW which is driven by lower metallicities as well as more recently-declining star formation histories relative. However, in our sample, we find that the UV faint (M$_{\rm{UV}} >-19$) galaxies have a higher H$\alpha$ EW which indicates lower metallicities and high star formation rates. The difference in trends compared to \cite{Endsley2023a} could arise from differences in sample selection strategies, where we have selected galaxies with F410M band excess that biases our sample towards star forming galaxies with higher nebular emission, whereas \cite{Endsley2023a} follow a Lyman-break selection which allows them to select galaxies that are not biased towards higher emission.  



\section{Conclusions}
In this work, we present the UV and ionisation properties and star formation of dwarf galaxies at z $\sim 5$ to 7. We use the CANUCS NIRCam and NIRSpec observations of the MACS0417 CLU field and the respective NCF field. Our sample is selected with a F410M band flux excess at $4.8<\rm{z}<5.5$ such that the H$\alpha$ emission is included in the medium band and we can measure accurate emission line flux and EWs. We confirm the line fluxes for a subset of the NIRCam sample with NIRSpec follow-up (figure \ref{fig:spec_phot}). With the magnification from the foreground cluster scale lens in our field, we are also able to push to fainter and lower mass galaxies compared to previous studies at similar redshifts.  Using this data, we discussed the evolution of \buv and \xiion and their dependence on the galaxy properties and star formation histories. We also study the ionising properties of our sample using emission line EWs. We summarise our results as follows:

1. We find bluer UV slopes of dwarf galaxies with median \buv  $-2.56 \pm 0.26$ in the NIRCam sample. We find no significant evolution of \buv with redshift during the epoch of reionisation but find a significant correlation between \buv and the stellar mass of the galaxy in both our sample sets. These results indicate a lack of dust build-up or a higher presence of ionising radiation in star forming dwarf galaxies at z$\sim5$. 

2. We do not find any evolution of \xiion with redshift in the EoR and no significant difference between \xiion of Ly$\alpha$ emitters reported in the literature with our sample (median $\log$ \xiion $=25.36\pm0.04$ and  $25.49\pm0.18$ in NIRCam and NIRSpec samples respectively, figure \ref{fig:xion_z}). The median $\log$ \xiion of our sample sets is also above the canonical value of 25.2  required to reionise the local IGM with approximately $20\%$ escape fraction \citep{Robertson2015}. The lack of difference between the median \xiion Ly$\alpha$ emitters and non-emitters indicates different physical processes affecting the production of ionising photons and the creation of escape channels for Lyman $\alpha$ photons. 

3. We find a moderate correlation of \xiion with the stellar mass and the specific star formation rates in a galaxy, such that \xiion decreases with increasing stellar mass and increases with increasing sSFR (figure \ref{fig:sm}). On further investigation of the star formation histories, we find that the majority of our H$\alpha$ selected galaxies are undergoing recent star formation raise/burst which creates a large scatter in the \xiion. We also find that the relative time of burst is correlated to average \xiion and indicates a delay in of $\approx 50$ Myrs between the final rise in star formation/burst and increasing \xiion (figure \ref{fig:sfh}). 

4. We investigate the correlation between \xiion and UV properties \buv and M$_{UV}$. We find a significant correlation between M$_{UV}$ and \xiion such that \xiion increases with decreasing M$_{UV}$. This implies a possible greater contribution of UV faint and dwarf galaxies to reionising the Universe. However, we find no significant and opposite correlation between \buv and \xiion. The NIRCam sample, which covers a small dynamic range of beta finds higher \xiion for redder \buv. The large scatter in \buv and \xiion which arise due to the stochastic nature of star formation histories in dwarf galaxies could explain the observed discrepancy. 

5. We also find a significant correlation between \xiion and EWs of restframe optical emission lines. We find that \xiion increases highly with increasing H$\alpha$ EW and moderately with [OIII]+ H$\beta$ EW. We further explore the correlation of restframe optical emission EWs with M$_{\rm{UV}}$. We find that on average H$\alpha$ EW increases with M$_{\rm{UV}}$, although our bright end of the  M$_{\rm{UV}}$ range has a high scatter in H$\alpha$ EW indicating a stochastic star formation history with very recent bursts. However, our UV faint sample shows a high H$\alpha$ EW. On the other hand, we do not find a significant correlation between [OIII]+ H$\beta$ EW and M$_{\rm{UV}}$. However, the lack of medium band observation covering the [OIII]+ H$\beta$ emission in the CLU field limits us to M$_{\rm{UV}} <-18$, where we similarly find a large scatter in [OIII]+ H$\beta$ EW.

Our results from one of the five cluster fields of CANCUS provide a large set of empirical measurements of UV and ionising properties of dwarf star forming galaxies at z $>5$. We demonstrate here the use of medium band photometry to accurately derive emission line fluxes without having to rely solely on spectroscopy. Similar studies characterising the properties of dwarf galaxies across the epoch of reionisation are required to fully understand how reionisation of the Universe progressed. This work will be supplemented by future studies with the full CANUCS data at z$>5$, which will provide a better understanding of the general population of galaxies across the history of the Universe.

\section*{Acknowledgements}

AH, MB, GR, and NM acknowledge support from the ERC Grant FIRSTLIGHT and Slovenian national research agency ARRS through grants N1-0238 and P1-0188. MB  acknowledges support from the program HST-GO-16667, provided through a grant from the STScI under NASA contract NAS5-26555.
This research was enabled by grant 18JWST-GTO1 from the Canadian Space Agency and funding from the Natural Sciences and Engineering Research Council of Canada. This research used the Canadian Advanced Network For Astronomy Research (CANFAR) operated in partnership by the Canadian Astronomy Data Centre and The Digital Research Alliance of Canada with support from the National Research Council of Canada the Canadian Space Agency, CANARIE and the Canadian Foundation for Innovation. The Cosmic Dawn Center (DAWN) is funded by the Danish National Research Foundation under grant No. 140. 
\section*{Data Availability}

Raw JWST data used in this work is available from the Mikulski
Archive for Space Telescopes (https://archive.stsci.edu; doi: 10.17909/ph4n-6n76).
Processed data products will be available on http://canucs-jwst.com.



\bibliographystyle{mnras}

\bsp	
\label{lastpage}
\end{document}